\documentstyle[12pt,epsfig]{article}
\newcommand{\be}[1]{\begin{equation} \label{(#1)}}
\newcommand{\ee}{\end{equation}}
\newcommand{\ba}[1]{\begin{eqnarray} \label{(#1)}}
\newcommand{\ea}{\end{eqnarray}}
\newcommand{\nn}{\nonumber}

\def\p{\prime}
\def\pmb#1{\setbox0=\hbox{#1}  \kern-.015em\copy0\kern-\wd0
  \kern.03em\copy0\kern-\wd0
  \kern-.015em\raise.0233em\box0 }

\def\rp{$R_p\hspace{-1em}/\ \ $}
\def\Lv{$L\hspace{-0.5em}/\ \ $}
\def\Lfv{$L_i\hspace{-0.8em}/\ \ $}
\def\Bv{$B\hspace{-0.6em}/\ \ $}
\def\rpm{R_p \hspace{-0.8em}/\;\:}

\def\lg{\langle}
\def\rg{\rangle}

\begin{document}
\begin{titlepage}

\begin{center}
%
{\Large\bf Exotic $\mu^--e^-$ conversion in nuclei and R-parity violating
supersymmetry}\\[1cm]

\bigskip

{Amand Faessler$^a$, T.S. Kosmas$^b$, Sergey Kovalenko$^{c,}$\footnote{On
leave of absence from the Joint Institut for Nuclear Research, Dubna, Russia}
and J.D. Vergados$^b$} \\[0.5cm]
{$^a$\it Institut f\"ur Theoretische Physik  der Universit\"at T\"ubingen, 
D-72076 T\"ubingen, Germany}\\[3mm]
{$^b$\it Division of Theoretical Physics, University of Ioannina GR-45110
Ioannina, Greece}\\[3mm]
{$^c$ \it Departamento de F\'\i sica, Universidad
T\'ecnica Federico Santa Mar\'\i a, Casilla 110-V, Valpara\'\i so, Chile}
\end{center}

\bigskip

\begin{abstract}
The flavor violating $\mu^--e^-$ conversion in nuclei is studied within
the minimal supersymmetric standard model. We focus on the R-parity
violating contributions at tree level including the trilinear and the
bilinear terms in the superpotential as well as in the soft supersymmetry
breaking sector. The nucleon and nuclear structure have consistently been
taken into account in the expression of the $\mu^--e^-$ conversion
branching ratio constructed in this framework. We have found that the
contribution of the strange quark sea of the nucleon is comparable with
that of the valence quarks.  From the available experimental data on
$\mu^--e^-$ conversion in $^{48}$Ti and $^{208}$Pb and the expected
sensitivity of the MECO experiment for $^{27}$Al we have extracted new
stringent limits on the R-parity violating parameters.
\end{abstract}

\bigskip
\bigskip


\bigskip
\bigskip


\end{titlepage}
\bigskip
\section{Introduction}

The lepton flavor violating process of neutrinoless
muon-to-electron ($\mu-e$) conversion in a nucleus, represented by
\begin{equation}
\mu^- + (A,Z) \longrightarrow  e^- \,+\,(A,Z)^*\, ,
\label{I.1}
\end{equation}
is an exotic process very sensitive to a plethora of new-physics
extensions of the standard model(SM)
\cite{Marci}-\cite{KLV94}.
In addition, experimentally
it is accessible with incomparable sensitivity. Long time ago Marciano and Sanda
\cite{Marci} has proposed it as one of the best probes to search for lepton
flavor violation beyond the standard model.
Recently, in view of the indications for neutrino oscillations
in super-Kamiokande,
solar neutrino
and LSND data, new hope has revived among the experimentalists of nuclear and
particle physics to detect other signals for physics beyond the
SM.
A prominent probe in this spirit is this exotic process (1).

The fact that the upper limits on the branching ratio of the $\mu^--e^-$
conversion relative to the ordinary muon capture,
\begin{eqnarray}\label{rat}
R_{\mu e^-}=\Gamma(\mu^-\to e^-)/\Gamma(\mu^-\to\nu_\mu),
\end{eqnarray}
offer the lowest constraints compared to any purely leptonic rare process
motivated a new $\mu^--e^-$ conversion experiment, the so called MECO experiment
at Brookhaven \cite{Molzon,Molz99,SSK99},
which got recently scientific approval and is planned to start soon.
The MECO experiment is going to use a new very intense $\mu^-$ beam and
a new detector operating at the Alternating Gradient Synchrotron (AGS). The
basic feature of this experiment is the use of a pulsed $\mu^-$ beam to
significantly reduce the prompt background from $\pi^-$ and $e^-$
contaminations. For technical reasons the MECO target has been chosen to be the
light nucleus $^{27}$Al.
Traditionally the $\mu - e$ conversion process was searched by employing medium heavy
(like $^{48}$Ti and $^{63}$Cu) \cite{Schaaf,Dohmen} or very heavy (like $^{208}$Pb
and $^{197}$Au) \cite{Schaaf,Honec,Ahmad} targets (for a historical review on such
experiments see Ref. \cite{KVF98}).
The best upper limits on $R_{\mu e^-}$ set up to the present have been extracted at
PSI by the SINDRUM II experiments resulting in the values
\begin{eqnarray}\label{Ti}
&&R_{\mu e^-} \leq 6.1\times 10^{-13}
\quad\mbox{ for }^{48}\mbox{Ti target \cite{Schaaf}},\\
\label{Pb}
&&R_{\mu e^-} \leq 4.6\times 10^{-11}
\quad\mbox{ for }^{208}\mbox{Pb target \cite{Dohmen}}.
\end{eqnarray}
(at 90\% confidence level).
The experimental sensitivity of the Brookhaven experiment is expected to be
roughly
\begin{eqnarray}\label{Al}
R_{\mu e^-} \leq 2\times 10^{-17}
\quad\mbox{for }^{27}\mbox{Al target \cite{Molzon,Molz99}}
\end{eqnarray}
i.e. three to four orders of magnitude below the existing experimental limits.

It is well known that the process (1) is a very good example of
the interplay between particle and nuclear physics attracting
significant efforts from both sides.
The underlying nuclear physics of the $\mu-e$ conversion has been
comprehensively studied in Refs.
\cite{KVF98}-\cite{KosmasVergados96}.
>From the particle physics point of view, processes like
$\mu-e$ conversion, is forbidden in the SM
by muon and electron quantum number conservation. Therefore
it has long been recognized as an important probe of the flavor changing
neutral currents and possible physics beyond the SM
\cite{Marci}-\cite{KLV94}.

On the particle physics side there are many mechanisms of
the $\mu-e$ conversion constructed in the literature (see
\cite{KLV94,9701381,Huitu,Kos-Kov}
and references therein). All these mechanisms fall into two
different categories:
photonic and non-photonic. Mechanisms from different categories significantly
differ from the point of view of the nucleon and nuclear structure
calculations. This stems from the fact that they proceed at different
distances and, therefore, involve different details of the structure.
The long-distance photonic mechanisms are mediated
by the photon exchange between the quark and the $\mu-e$-lepton currents.
These mechanisms resort to the lepton-flavor non-diagonal
electromagnetic vertex which is presumably induced by
the non-standard model physics at the loop level.
The contributions to the $\mu-e$ conversion via virtual
photon exchange exist in all models which allow $\mu\rightarrow e \gamma$
decay.
The short-distance non-photonic mechanisms contain heavy
particles in intermediate states and can be realized at
the tree level,  at the 1-loop level or at the level of box diagrams.

The non-photonic mechanisms are mediated by various particles in intermediate states
such as $W,Z$-bosons
\cite{Marci}-\cite{KLV94}, Higgs bosons \cite{Shanker,KLV94},
supersymmetric particles (squarks, sleptons, gauginos, higgsinos etc.)
with and without $R$-parity conservation in the vertices.

In the supersymmetric (SUSY) extensions of the SM with conserved R-parity
($R_p$SUSY) $\mu^--e^-$ conversion has been studied in Refs.
\cite{KosmasVergados96,R-cons}.  In this case the SUSY contributions
appear only at the loop or box level and, therefore, they are suppressed
by the loop factor. The situation is different in the SUSY models with
R-parity non-conservation (\rp SUSY). In this framework there exist
the tree level non-photonic contributions
\cite{9701381,Kos-Kov} and the 1-loop photonic contributions
significantly enhanced by the large logarithms \cite{Huitu}.

The primary purpose of this work is to offer a theoretical background
for the running and planned $\mu-e$ conversion experiments. We consider
all the possible tree level contributions to $\mu-e$ conversion in the
framework of the minimal SUSY model with most
general form of R-parity violation (\rp  MSSM) including the trilinear
\rp couplings and the bilinear \rp lepton-Higgs terms.
We also examine some non-SUSY and $R_p$SUSY
mechanisms previously studied in \cite{mu-e-nucl,KosmasVergados96,R-cons}.
We develop a formalism of calculating the $\mu-e$ conversion rate for the
quark level Lagrangian with all these terms.

In our study we pay special attention to the effect of nucleon and nuclear
structure dependence of the $\mu-e\ $ conversion branching ratio $R_{\mu e^-}$.
In particular, we take into account the contribution of
the strange nucleon sea which, as we will see, gives a
contribution comparable to the contribution of the valence
quarks of a nucleon.

Thus, we apply our formalism to the case of nuclei $^{48}Ti$ and $^{208}Pb$
by calculating numerically the muon-nucleus overlap integral and
solving the Dirac equations with modern neural networks techniques and
using the PSI experimental data. A similar application is done for the
$^{27}$Al target by employing the sensitivity of the designed MECO experiment.

Our final goal is to derive on this theoretical basis
the experimental constraints on the $R_{p}\hspace{-1em}/$  Yukawa couplings,
the lepton-Higgs mixing parameters and on the sneutrino VEVs.
Towards this end we use the experimental upper limits on
the branching ratio $R_{\mu e^-}$ given above and derive
the new stringent constraints on the R-parity
violating parameters.

\section{The effective $\mu - e$ conversion Lagrangian }

It is well known that, the lepton flavors $(L_i)$ and the total
baryon $(B)$ number are conserved by the standard model interactions
in all orders of perturbation theory. As mentioned above, this is
an accidental consequence of the SM field content and gauge invariance.
Thus $\mu-e$ conversion is forbidden in the SM.
In contrast to $L_i$ the individual quark flavors $B_i$ are not conserved
in the SM by the charged current interactions due to the presence of
non-trivial Cabibbo-Kobayashi-Maskawa (CKM) mixing matrix.
$L_i$ are conserved since the analogous mixing matrix can be rotated
away by the neutrino fields redefinition.
The latter is possible while the neutrino fields
have no mass term and, therefore, are defined up to an arbitrary unitary
rotation.
As soon as the model is extended by inclusion of the right-handed
neutrinos lepton flavor violation can occur since neutrinos may
acquire a non-trivial mass matrix.

In the supersymmetric extensions of SM $L_i$ and $B$ conservation
laws are in general violated. As a result potentially dangerous
total lepton (\Lv) and baryon (\Bv) number violating processes become open.

One may easily eliminate
\Lv and \Bv interactions from a SUSY model by
introducing a discrete symmetry known as R-parity. This is a multiplicative
$Z_{2}$ symmetry defined as $R_{p}=(-1)^{3B+L+2S}$, where $S$ is the spin
quantum number. In this framework neutrinos are massless.
However the flavor violation in the lepton sector can occur
at the 1-loop level via the $L_i$-violation in the slepton sector.
Thus, $\mu^--e^-$ conversion is allowed in the SUSY models with R-parity
conservation.

There is no as yet convincing theoretical motivation
for R-symmetry of the low energy Lagrangian and, therefore, SUSY models
with ($R_p$SUSY) and without ($R_{p}\hspace{-1em}/\ \ $ SUSY) R-parity
conservation are "a priori" both plausible.

Despite the above mentioned problems with \Lv, \Bv interactions
$R_{p}\hspace{-1em}/\ \ $ SUSY looks rather attractive, since
it may offer a clue to the solution of some long standing problems of
particle physics, such as neutrino mass problem.
In the \rp  SUSY framework neutrinos acquire Majorana masses at the weak-scale
via mixing with the gauginos and higgsinos as well as via \Lv loop effects
\cite{neutr}. Furthermore, \rp  SUSY models admit non-trivial contributions to the
lepton flavor violating processes. During the last few years the \rp  SUSY
models have been extensively studied in the literature (for a recent review see
Refs. \cite{reviews}).

We analyze possible mechanisms for process (1) existing at the tree level in
the minimal \rp SUSY model with a most general form of R-parity violation.

A most general gauge invariant form of
the R-parity violating part of the superpotential
at the level of renormalizable operators reads
\begin{eqnarray}
W_{\rpm} =\lambda _{ijk}L_{i}L_{j}E_{k}^{c} + \bar\lambda _{ijk}^{\prime
}L_{i}Q_{j}D_{k}^{c}+\mu _{j}L_{j}H_{2}+\bar\lambda _{ijk}^{\prime \prime
}U_{i}^{c}D_{j}^{c}D_{k}^{c},
\label{sup1}
\end{eqnarray}
where $L$, $Q$ stand for lepton and quark doublet left-handed superfields
while $E^{c},\ U^{c},\ D^{c}$ for lepton and {\em up}, {\em down} quark
singlet superfields; $H_{1}$ and $H_{2}$ are the Higgs doublet superfields
with a weak hypercharge $Y=-1,\ +1$, respectively. Summation over the
generations is implied. The coupling constants
$\lambda $ ($\bar\lambda ^{\prime\prime }$) are antisymmetric
in the first (last) two indices.
The bar sign in $\bar\lambda', \bar\lambda'' $ denotes
that all the definitions are given in the gauge basis for
the quark fields. Later on we will change to the mass basis
and drop the bars.
Henceforth we set $\bar\lambda^{\prime \prime }=0$ which are
irrelevant for our consideration. This condition ensures the proton
stability and can be guaranteed by special discreet symmetries other
than R-parity.

The R-parity non-conservation brings into the SUSY phenomenology
the lepton number (\Lv) and lepton flavor (\Lfv) violation originating
from the two different sources.
One is given by the \Lv trilinear couplings in
the superpotential $W_{R_{p}\hspace{-0.8em}/\;}$ of Eq. (\ref{sup1}).
Another is related to the bilinear terms in $W_{R_{p}\hspace{-0.8em}/\;}$
and in soft SUSY breaking sector.
Presence of these bilinear terms leads to the terms linear in the sneutrino
fields $\tilde{\nu}_{i}$ in the scalar potential. The linear terms drive these
fields to non-zero vacuum expectation values $\langle \tilde{\nu}_{i}\rangle\neq 0$
at the minimum of the scalar potential.
At this ground state the MSSM vertices $\tilde{Z}\nu$ $ \tilde{\nu}$ and
$\tilde{W}e\tilde{\nu}$ produce the gaugino-lepton mixing mass terms
$\tilde{Z}\nu\langle \tilde{\nu}\rangle ,\tilde{W}e\langle \tilde{\nu}\rangle $
(with $\tilde{W},\tilde{Z}$ being wino and zino fields). These terms
taken along with the lepton-higgsino $\mu _{i}L_{i}\tilde{H}_{1}$ mixing
from the superpotential of Eq. (\ref{sup1})
form $7\times 7$ neutral fermion
and $5\times 5$ charged fermion mass matrices.
For the considered case of $\mu-e$ conversion the only charged
fermion mixing is essential.
The charged fermion mass term takes the form
\begin{eqnarray}\label{m_term}
{\cal L}^{(\pm)}_{mass} =
- \Psi_{(-)}^{\p T} {\cal M}_{\pm} \Psi_{(+)}^{\p} -\mbox{H.c.}
\end{eqnarray}
in the basis of  two component Weyl spinors corresponding to
the weak eigenstate fields
\begin{eqnarray}
\Psi _{(-)}^{\prime T} =(e_{\ L}^{-},\,\,\mu _{\ L}^{-},\,\,
\tau _{\ L}^{-},\,\,-i\lambda _{-},\,\,\tilde{H}_{1}^{-}), \ \ \
\Psi _{(+)}^{\prime T} &=&(e_{\ L}^{+},\,\,\mu _{\ L}^{+},\,\,\tau _{\
L}^{+},\,\,-i\lambda _{+},\,\,\tilde{H}_{2}^{+}).
\end{eqnarray}
Here $\lambda _{\pm}$ are the $SU_{2L}$ gauginos while the higgsinos are
denoted as $\tilde{H}_{1,2}^{\pm }$.
These fields are related to the mass eigenstate fields $\Psi _{(\pm )}$
by the rotation
\begin{equation}
\Psi _{(\pm )i}=\Delta _{ij}^{\pm }\ \Psi _{(\pm )j}^{\prime },  \label{mix}
\end{equation}
The unitary mixing matrices $\Delta ^{\pm }$ diagonalize the chargino-charged
lepton mass matrix as
\begin{eqnarray}\label{diag+}
\left(\Delta^{-}\right)^*{\cal M}_{\pm}\left(\Delta^{+}\right)^{\dagger} =
Diag\{m^{(l)}_i, m_{\chi^{\pm}_k}\},
\end{eqnarray}
where $m^{(l)}_i$ and  $m_{\chi^{\pm}_k}$ are the physical charged
lepton and chargino masses. In the present paper we use the notations
of Refs. \cite{Now,FKS97}.

Rotating the MSSM Lagrangian to the mass eigenstate
basis according to Eq. (\ref{mix}) one obtains the new lepton
number and lepton flavor violating interactions in addition to those
which are present in the superpotential Eq. (\ref{sup1}).
Note that the mixing between the charged
leptons $(e_{\ L}^{+},\,\,\mu _{\ L}^{+})$
and the chargino components $(-i\lambda_{+},\,\, \tilde H^+_2)$,
described by the off diagonal blocks of the $\Delta^+$,
is proportional to the small factor $\sim m_{e,\mu}/M_{SUSY}$
\cite{Now,FKS97} and is, therefore, neglected in our analysis.

Let us write down the MSSM terms generating by the rotation (\ref{mix})
the new lepton flavor violating interactions relevant for the $\mu^--e^-$
conversion. In the two component form they can be written as \cite{mssm}
\begin{eqnarray}\label{MSSM-rel}
{\cal L}_{_{MSSM}} = \frac{g_2}{2 \cos\theta_W} Z^{\mu}
\bar{\Psi'}_i A_{ij} \bar\sigma_{\mu} \Psi'_j + i g_2 \lambda^- u_L
\tilde d^*_L,
\end{eqnarray}
where $A_{ij} = (1 -2 \sin^2\theta_W)\delta_{ij} + \delta_{i4}\delta_{4j}$.
Rotating this equation to the mass eigenstate basis
we write down in the four-component Dirac notation
\begin{eqnarray}\label{MSSM-mes}
{\cal L}_{_{MSSM}} = \frac{g_2}{2 \cos\theta_W} a_Z Z^{\mu}
\bar e \gamma_{\mu} P_{_L} \mu + g_2 \zeta_i \cdot
\bar u_k P_{_R} e^c_i \tilde d_{Lk}.
\end{eqnarray}
Here $e_{i} = (e, \mu, \tau)$,  $P_{_{L,R}}=(1\mp \gamma_{5})/2$.
The lepton flavor violating parameters in this formula are given by
\begin{eqnarray}\label{param1}
a_Z = \Delta^-_{14} {\Delta^{-}}^*_{24} \approx \xi_{11} \xi^*_{21},
\ \ \ \ \ \ \zeta_i = {\Delta^{-}}^*_{i4} \approx \xi_{i1}.
\end{eqnarray}
The approximate expressions for these parameters were found
by the method of Ref. \cite{Now} expanding the mixing matrix
$\Delta^-$ in the small matrix parameter
\begin{eqnarray}\label{xi-mix}
\xi_{i1}^{*} = \frac{g_{2}(\mu \lg\tilde\nu_i\rg - \lg H_1\rg \mu_i)}{\sqrt{2}\
(M_{2}\mu -\sin \!2\beta M_{W}^{2})},
\end{eqnarray}
where $\mu, M_2$ are the ordinary MSSM mass parameters from
the superpotential term $\mu H_1 H_2$ and from the $SU_2$ gaugino
soft mass term $M_2 \tilde W^k \tilde W^k$. The MSSM mixing angle
is defined as $\tan\beta = \lg H_2\rg/\lg H_1\rg$.
The other lepton flavor violating interaction contributing to the
$\mu^--e^-$ conversion come from the superpotential (\ref{sup1}).

The leading diagrams describing possible tree
level \rp  MSSM contributions to the $\mu-e$ conversion are
presented in Fig. 1. The vertex operators encountered in these
diagrams are
\begin{eqnarray}\nn
&&{\cal L}_{\mu e^-} =
2\,\lambda _{i21}\,\,\tilde{\nu}_{_{Li}}
\bar{e}\,P_{_L}\,\,\mu +
\lambda _{ijj}^{\prime }\,\tilde{\nu}_{_{Li}} \bar{d}_{_j}\,P_{_L}\,d_{_j} -
\lambda_{ijk}^{\prime *}V_{nj}\left(\tilde{u}_{_{Ln}}^{*}\overline{e_i}
\,P_{_R}\,d_k + \tilde{d}_{_{Rk}}\bar{u}_{_n} \,P_{_R}\,e^{c}_i\right) +\\
&&+ g_{2}\,\zeta_{i}\ V_{nk}\cdot \bar{u}_n P_{_R}e_{_i}^{c}\ \tilde{d}_{_{Lk}}+
\frac{1}{2} \frac{g_2}{\cos\theta_W} a_{Z}
Z^{\mu }\bar{e}\gamma _{\mu }P_{_L}\mu -
\frac{g_2}{\cos\theta_W} Z^{\mu}
\bar q\gamma_{\mu}(\epsilon_L P_{_L} + \epsilon_R P_{_R})q,
\label{q-lev}
\end{eqnarray}
where the first three terms originate from the superpotential
(\ref{sup1}), the fourth and fifth terms correspond to the chargino-charged
lepton mixing terms in Eq. (\ref{MSSM-mes}) and the last one is the ordinary
SM neutral current interaction.

\begin{figure}[h!]
\vspace{-3.cm}
\hspace{-2.58cm}
\mbox{\epsfxsize=19. cm\epsffile{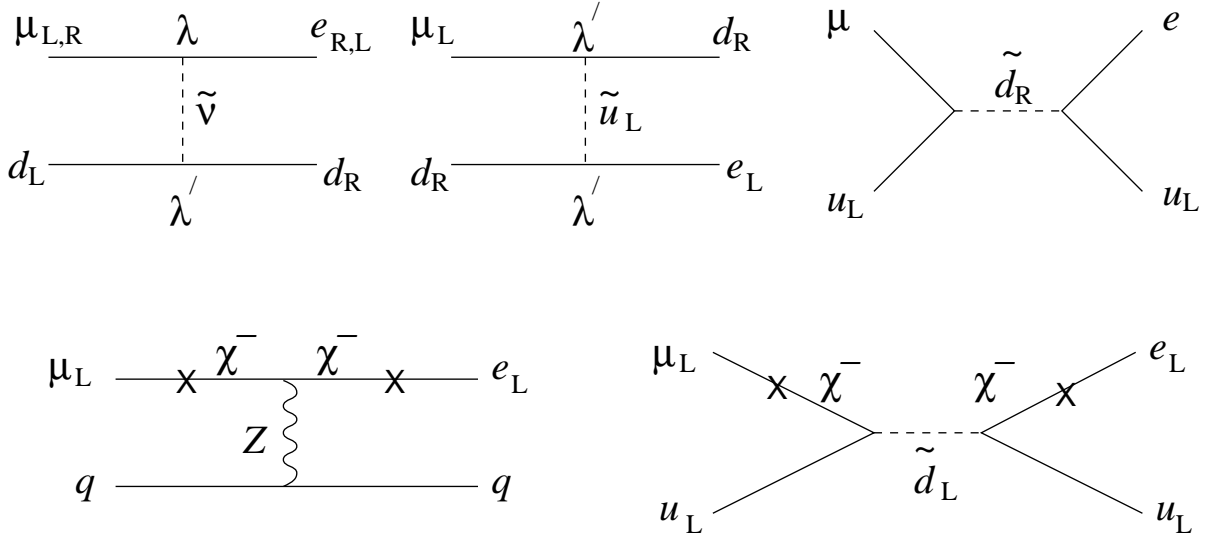}}
\vspace{-14cm}
\caption{
Leading \rp~MSSM diagrams contributing to $\mu-e$ conversion at the tree level.
(i) The upper diagrams originate from the trilinear $\lambda$, $\lambda'$
terms in the superpotential Eq. (\ref{sup1}).
(ii) The lower diagrams originate from the chargino-charged lepton mixing
schematically denoted by crosses (X) on the lepton lines.}  
\vskip 1cm  
\end{figure}

In Eq. (\ref{q-lev}) the SM neutral current parameters
are defined as usual
$$
\epsilon_L = T_3 - \sin^2\theta_W \ Q \ , \qquad
\epsilon_R =  - \sin^2\theta_W \ Q
$$
with $T_3$ and $Q$ being the 3rd component of the weak isospin and
the electric charge.

The Lagrangian (\ref{q-lev}) is given in the quark mass eigenstate basis
which is related to the flavor basis $q^0$ through
$$q_{L,R} = V_{L,R}^q \cdot q^0.$$
For convenience we introduced the new couplings
$$\lambda_{ijk}^{\prime} = \bar\lambda_{imn}^{\prime }
\left(V^d_L\right)^*_{jm} \left(V^d_R\right)_{kn}.$$
The CKM matrix
is defined in the standard way as  $V = V^u_L V^{d\dagger}_L.$

Integrating out the heavy fields from the diagrams in Fig. 1
and carrying out Fierz reshuffling we obtain the 4-fermion effective
Lagrangian which describes the $\mu-e$ conversion at the quark level
in the first order of perturbation theory. It takes the form
\begin{equation}
{\cal L}_{eff}^{q}\ =\ \frac{G_F}{\sqrt{2}}j_{\mu }\left[ \eta _{L}^{ui}J_{uL(i)}^{\mu }+
\eta_{R}^{ui}J_{uR(i)}^{\mu }+\eta_{L}^{di}J_{dL(i)}^{\mu }+
\eta_{R}^{di}J_{dR(i)}^{\mu}\right] + \frac{G_F}{\sqrt{2}}\left[
\bar\eta_{R}^{di}J_{dR(i)}j_{L} + \bar\eta_{L}^{di}J_{dL(i)}j_{R}\right].
\label{eff-q}
\end{equation}
The index $i$ denotes generation so that $u_i = u,c,t$ and $d_i = d,s,b$.
The coefficients $\eta $ accumulate dependence on the $R_{p}\hspace{-1em}/\ \ $
SUSY parameters in the form
\begin{eqnarray} \nn
\eta_{L}^{ui} &=&- \frac{1}{\sqrt{2}} \sum_{l,m,n}\frac{\lambda _{2ln}^{\prime }
\lambda_{1mn}^{\prime *}}{G_F \tilde m_{dR(n)}^{2}}V^*_{il} V_{im} +
\,2(1-\frac{4}{3}\sin^{2}\theta _{W})\, a_Z +
4 \sum_n \, {\zeta_{1} \zeta_{2}^*}
\frac{M_W^2}{\tilde m_{dL(n)}^2}|V_{in}|^2,\\
\label{coeff12}
\eta_{R}^{di} &=&  \frac{1}{\sqrt{2}}\sum_{l,m,n}
\frac{\lambda_{2mi}^{\prime }\lambda_{1li}^{\prime *}}
{G_F \tilde m_{{u}L(n)}^{2}} V^*_{nm} V_{nl}
+\frac{4}{3} \sin^{2}\theta _{W} \, a_Z,  \\ \nn
\eta_{L}^{di} &=& - 2(1-\frac{2}{3} \sin^{2}\theta_{W}) \, a_Z, \  \  \
\eta_{R}^{ui} =
- \frac{8}{3} \sin^{2}\theta_{W} \, a_Z
\\ \nn
\bar\eta_{L}^{di} &=& - \sqrt{2} \sum_{n}
\frac{\lambda_{nii}^{\prime }\lambda _{n12}^{*}}{{G_F \tilde m^2_{\nu(n)}}},
\ \ \
\bar\eta_{R}^{di} = - \sqrt{2} \sum_{n}
\frac{\lambda_{nii}^{\prime*}\lambda_{n21}}{G_F \tilde m^2_{{\nu}(n)}}.
\end{eqnarray}
Here $\tilde m_{q(n)}, \tilde m_{\nu(n)}$ are the squark and sneutrino
masses.
In Eq. (\ref{eff-q}) we introduced the color singlet currents
\begin{eqnarray}\label{currents}
J_{q_{L/R}(i)}^{\mu }=\bar{q}_i \gamma^{\mu }P_{_{L/R}}q_{i},\ \ \
J_{d_{L/R}(i)}=\bar{d}_i P_{_{L/R}}d_{i}, \ \ \
j^{\mu }=\bar{e}\gamma^{\mu} \mu, \ \ \
j_{_{L/R}}=\bar{e} P_{_{L/R}}\mu,
\end{eqnarray}
where $q_i = (u_i, d_i)$.

Since in the next sections we report the new results for
the nuclear matrix elements of $^{48}$Ti, $^{27}$Al and $^{208}$Pb
we are going also to update $\mu^-\to e^-$ constraints on
the effective lepton flavor violating parameters corresponding
to certain non-SUSY and $R_p$SUSY mechanisms shown in Fig. 2.
\begin{figure}[h!]
 \vspace{-2cm}
 \vspace{-1.cm}
 \hspace{1.0cm}
 \mbox{\epsfxsize=16. cm\epsffile{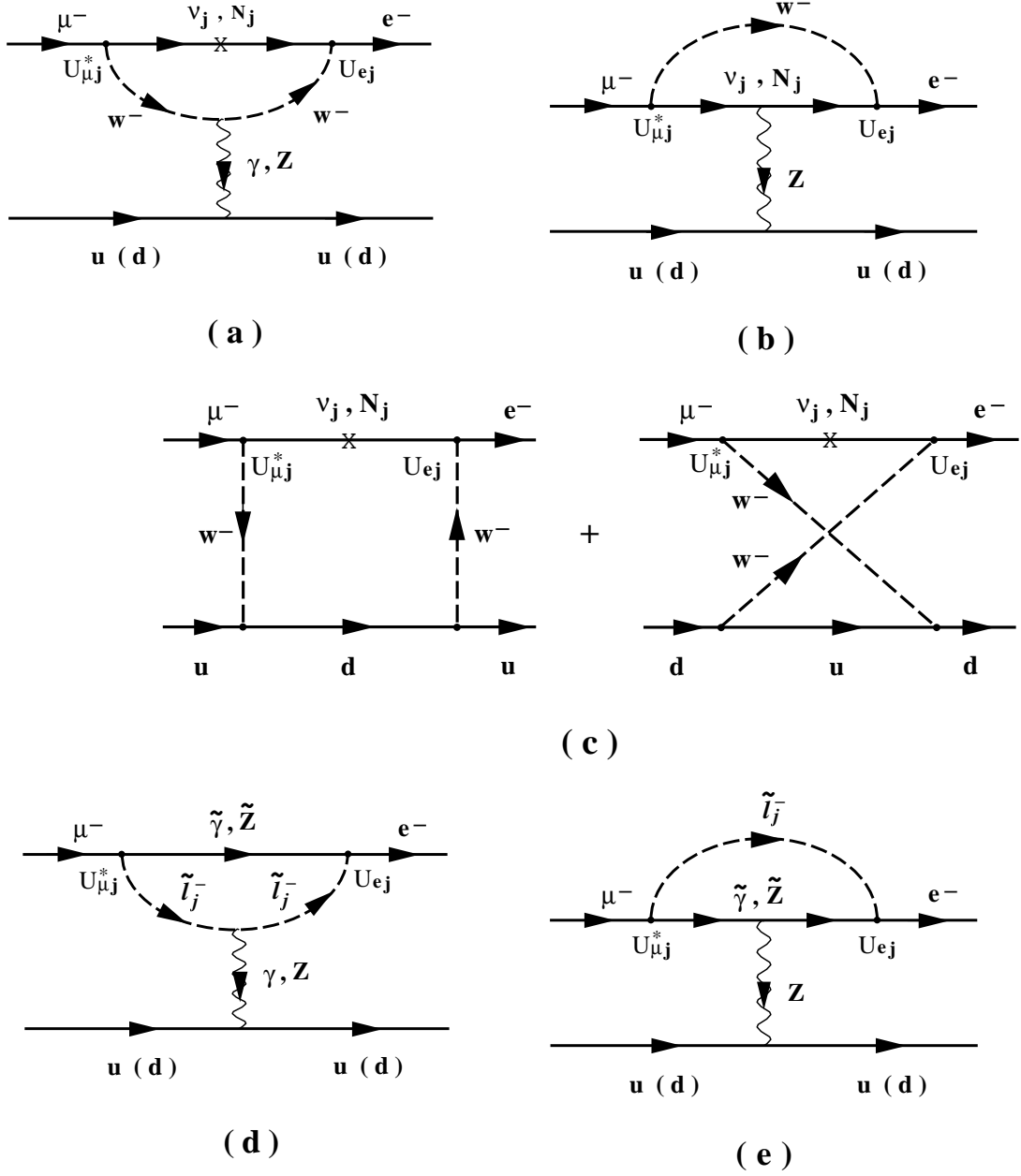}}
 \vspace{-3cm}
 \caption{Photonic and non-photonic mechanisms of the $\mu^-- e^-$ conversion within 
some extensions of the standard model: (a-c) the SM with massive neutrinos
and (a-c) R-parity conserving supersymmetric extensions of the SM.
}
\label{fig1}
\end{figure}                 
These mechanisms were previously studied in Refs. \cite{KosmasVergados96,R-cons}.
Let us shortly summarize these mechanisms for completeness.
The long-distance photonic mechanisms mediated
by the photon exchange between the quark and
the $\mu-e$-lepton currents is realized at the 1-loop level
as the $\nu-W$ loop [Fig. 2(a)] with the massive neutrinos
$\nu_i$ and the loop with the supersymmetric particles such as
the neutralino(chargino)-slepton(sneutrino) [Fig. 2(d)].
In the R-parity violating SUSY models there are also lepton-slepton
and quark-squark loops generated by the superpotential couplings
$\lambda L L E^c$ and $\lambda' L Q D^c$ respectively \cite{Huitu}.
The short-distance non-photonic mechanisms in
Fig. 2 contain heavy particles in intermediate states
and is realized at the 1-loop level [Fig. 2(a,b,d,e)]
or at the level of box diagrams [Fig. 2(c)].
The 1-loop diagrams of the non-photonic mechanisms
include the diagrams similar to those for the photonic
mechanisms but with the Z-boson instead of the photon [Fig. 2(a,d)]
as well as additional Z-boson couplings to the neutrinos and
neutralinos [Fig. 2(b,e)].
The box diagrams are constructed of the W-bosons and massive neutrinos
[Fig. 2(c)] as well as similar boxes with neutralinos and sleptons
or charginos and sneutrinos. The branching ratio formula for these mechanisms
is given in the next section.

\section{Nucleon and nuclear Structure dependence of the $\mu-e$ conversion rates.}

One of the main goals of this paper is the calculation of the $\mu-e$
conversion rate using realistic form factors of the participating nucleus
(A,Z).
This can be achieved by applying the conventional approach based on
the well known non-relativistic impulse approximation, i.e. treating
the nucleus
as a system of free nucleons \cite{Berna}. To follow this approach as a first
step one has to reformulate the $\mu-e$ conversion effective Lagrangian
(\ref{eff-q}) specified at the quark level in terms of the nucleon
degrees of freedom.

The transformation of the quark level effective Lagrangian, ${\cal L}_{eff}^{q}$,
to the effective Lagrangian at the nucleon level, ${\cal L}_{eff}^{N}$,
is usually done by utilizing the on-mass-shell matching condition \cite{FKSS97}
\begin{equation}
\langle \Psi_F|{\cal L}_{eff}^{q}|\Psi_I\rangle \approx \langle \Psi_F|{\cal L}%
_{eff}^{N}|\Psi_I\rangle ,  \label{match}
\end{equation}
where $|\Psi_I\rangle $ and $\langle \Psi_F|$ are the initial and
final nucleon states. In order to solve this equation we use
various relations for the matrix elements of the quark operators
between the nucleon states
\begin{eqnarray}\label{mat-el1}
\langle N|\bar{q}\ \Gamma_{K}\ q|N\rangle = G_{K}^{(q,N)}
\bar{\Psi}_N\ \Gamma_{K}\ \Psi_N,
\end{eqnarray}
with $q=\{u,d,s\}$,  $N=\{p,n\}$ and  $K = \{V,A,S,P\}$,
$\Gamma_K = \{\gamma_{\mu}, \gamma_{\mu}\gamma_5, 1, \gamma_5\}$.
Since the maximum momentum transfer ${\bf q}^{2}$ in $\mu -e$ conversion,
i.e. $|{\bf q}| \approx m_\mu/c$ with $m_\mu=105.6 MeV$ the muon mass,
is much smaller than the typical scale of nucleon structure we can safely neglect
the ${\bf q}^{2}$-dependence of the nucleon form factors $G_{K}^{(q,N)}$.
For the same reason we drop the weak magnetism and
the induced pseudoscalar terms proportional to the small
momentum transfer.

The isospin symmetry requires that
\begin{eqnarray}\label{isosym}
G_{K}^{(u,n)}=G_{K}^{(d,p)}\equiv G_{K}^{d}, \ \ \
G_{K}^{(d,n)}=G_{K}^{(u,p)}\equiv G_{K}^{u},\ \ \
G_{K}^{(s,n)}=G_{K}^{(s,p)}\equiv G_{K}^{s},
\end{eqnarray}
with $K=V,A,S,P$.
Conservation of vector current postulates the vector charge to be equal to
the quark number of the nucleon. This allows fixing of the vector nucleon
constants
\begin{eqnarray}\label{cvc}
G_{V}^{u}=2,\ \ \  G_{V}^{d}=1, \ \ \ G_{V}^{s}=0.
\end{eqnarray}
The axial-vector form factors $G_A$ can be extracted from the experimental
data on polarized nucleon structure functions \cite{EMC,SMC} combined with
the data on hyperon semileptonic decays \cite{Hyp}. The result is
\begin{eqnarray}\label{axial}
G_A^{u} \approx 0.78,\ \ \  G_A^{d} \approx -0.47, \ \ \
G_A^{s} \approx -0.19.
\end{eqnarray}
The scalar form factors are extracted from the baryon octet $B$ mass
spectrum $M_B$ in combination with the data on the pion-nucleon sigma term
\begin{eqnarray}\label{sigma}
\sigma_{\pi N} = (1/2)(m_u + m_d)\lg p|\bar u u + \bar d d|p\rg.
\end{eqnarray}
Towards this end we follow the QCD picture of the baryon masses which
is based on the relation \cite{SVZ,Cheng}
\begin{eqnarray}\label{emt}
\lg B|\theta^{\mu}_{\mu}|B\rg = M_B \bar B B
\end{eqnarray}
and on the well known representation for the trace of
the energy-momentum tensor \cite{SVZ}
\begin{eqnarray}\label{trace}
\theta^{\mu}_{\mu} = m_u \bar u u + m_d \bar d d
+ m_s \bar s s - (\tilde b\alpha_s/8\pi)G_{\mu\nu}^a G^{\mu\nu}_a,
\end{eqnarray}
where $G_{\mu\nu}^a$ is the gluon field strength, $\alpha_s$ is
the QCD coupling constant and $\tilde b$ is the reduced Gell-Mann-Low
function with the heavy quark contribution subtracted. Using these
relations in combination with $SU(3)$ relations \cite{Cheng} for
the matrix elements $\lg B|\theta^{\mu}_{\mu}|B\rg$  as well as
the experimental data on $M_B$ and  $\sigma_{\pi N}$ we derive
\begin{eqnarray}\label{scalar}
G_S^{u} \approx 5.1, \ \ \ G_S^{d} \approx 4.3, \ \ \  G_S^{s} \approx 2.5.
\end{eqnarray}

The nucleon matrix elements of the pseudoscalar
quark currents can be related to the divergence of the baryon octet
axial vector currents \cite{Cheng}. Utilizing this fact we
find the pseudoscalar form factors
\begin{eqnarray}\label{pseud}
G_P^{u} \approx 103, \ \ \ G_P^{d} \approx 100, \ \ \ G_P^{s} \approx 3.3.
\end{eqnarray}

Note that the strange quarks of the nucleon
sea significantly contribute to the nucleon form factors
$G_A$, $G_P$ and $G_S$. This result dramatically differs from
the na\"{\i}ve quark model and the MIT bag model where $G_{A,S,P}^{s} = 0$.
The contribution of the strange nucleon sea will allow us to extract
additional constraints on the second generation \rp   parameters.

Now we can solve Eq. (\ref{match}) and write
the effective $\mu-e$ conversion Lagrangian at the nucleon level as
\begin{equation}
{\cal L}_{eff}^{N}= \frac{G_F}{\sqrt{2}}\left[\bar{e}
\gamma_{\mu }(1-\gamma _{5})\mu \cdot J^{\mu }+
\bar{e}\mu \cdot J^{+}+\bar{e}\gamma_{5}\mu \cdot J^{-}\right].  \label{nucl1}
\end{equation}
where we have introduced the nucleon currents
\begin{eqnarray}
J^{\mu } &=&\bar{N}\gamma ^{\mu }\left[ (\alpha _{V}^{(0)}+\alpha
_{V}^{(3)}\tau _{3})+(\alpha _{A}^{(0)}+\alpha _{A}^{(3)}\tau _{3})\gamma
_{5}\right] \,N,  \label{nucur} \\
J^{\pm } &=&\bar{N}\,\,\left[ (\alpha _{\pm S}^{(0)}+\alpha _{\pm
S}^{(3)}\tau _{3})+(\alpha _{\pm P}^{(0)}+\alpha _{\pm P}^{(3)}\tau
_{3})\gamma _{5}\right] \,N,  \nonumber
\end{eqnarray}
for nucleon isospin doublet $N^T = (p, n)$.
The coefficients in Eq. (\ref{nucur}) are defined as
\begin{eqnarray}
\alpha _{V}^{(0)} &=&\frac{1}{8}(G_{V}^{u}+
G_{V}^{d})(\eta _{R}^{u1}+\eta_{L}^{u1}+\eta _{R}^{d1}+\eta _{L}^{d1}),
\nonumber \\ \nn
\alpha _{V}^{(3)} &=& \frac{1}{8}%
(G_{V}^{u}-G_{V}^{d})(\eta_{R}^{u1}+\eta_{L}^{u1}-\eta_{R}^{d1}-
\eta_{L}^{d1}),  \nonumber \\ \nn
\alpha _{A}^{(0)} &=&\frac{1}{8}(G_{A}^{u}+
G_{A}^{d})(\eta _{R}^{u1}-\eta_{L}^{u1}+\eta_{R}^{d1}-\eta_{L}^{d1}) +
\frac{1}{4}G_A^{s}(\eta _{R}^{d2} - \eta _{L}^{d2}),\\ \nn
\alpha _{A}^{(3)} &=&
\frac{1}{8}(G_{A}^{u}-G_{A}^{d})(\eta_{R}^{u1}-\eta _{L}^{u1}-
\eta_{R}^{d1}+\eta_{L}^{d1}),
 \label{alp1} \\
\alpha_{\pm S}^{(0)} &=&\frac{1}{16}(G_{S}^{u}+G_{S}^{d})
(\bar\eta_{L}^{d1}\pm \bar\eta_{R}^{d1}) +
\frac{1}{8}G_S^{s}(\bar\eta_{L}^{d2} \pm \bar\eta_{R}^{d2}),\\
\alpha_{\pm S}^{(3)} &=& -\frac{1}{16}
(G_{S}^{u}-G_{S}^{d})(\bar\eta_{L}^{d1}\pm \bar\eta_{R}^{d1}),
 \nonumber \\ \nn
\alpha_{\pm P}^{(0)} &=&-\frac{1}{16}(G_{P}^{u}+G_{P}^{d})
(\bar\eta_{L}^{d1}\mp \bar\eta_{R}^{d1}) -
\frac{1}{8}G_P^{s}(\bar\eta_{L}^{d2} \mp \bar\eta_{R}^{d2}),
 \nonumber \\ \nn
\alpha_{\pm P}^{(3)} &=& \frac{1}{16}
(G_{P}^{u}-G_{P}^{d})(\bar\eta_{L}^{d1}\mp \bar\eta_{R}^{d1}),
\ \ \   \nonumber
\end{eqnarray}

Starting from the Lagrangian
(\ref{nucl1}) it is straightforward to deduce the formula
for the total $\mu-e$ conversion rate. In the present paper
we focus on the coherent process, i.e. ground state to ground
state transitions. This is a dominant channel of $\mu-e$ conversion which,
in most of the experimentally interesting nuclei, exhausts more than
$90\%$ of the total $\mu^- \to e^-$ branching ratio 
\cite{Chiang,mu-e-nucl}.
To the leading order of the non-relativistic expansion the coherent
$\mu-e$ conversion rate takes the form
\begin{equation}
\Gamma_{\mu e^-}^{coh} = \frac{G^2_F p_e E_e}
{2 \pi } \ {\cal Q} ({\cal M}_p + {\cal M}_n)^2 \, ,
\label{Rme}
\end{equation}
where
\begin{eqnarray}
{\cal Q} \,=\, 2|\alpha_V^{(0)}+\alpha_V^{(3)}\ \phi|^2 +
|\alpha_{+S}^{(0)}+\alpha_{+S}^{(3)}\ \phi|^2 + |\alpha_{-S}^{(0)}+
\alpha_{-S}^{(3)}\ \phi|^2
\nonumber  \\
+2\ {\rm Re}\{(\alpha_V^{(0)}+\alpha_V^{(3)}\ \phi)[\alpha_{+S}^{(0)}+\alpha_{-S}^{(0)}+
(\alpha_{+S}^{(3)}+\alpha_{-S}^{(3)})\ \phi] \}\, .
\label{Rme.1}
\end{eqnarray}
The transition matrix elements
${{\cal M}}_{p,n}$
in Eq. (\ref{Rme})
depend on the final nuclear state populated during the $\mu-e$ conversion.
We should stress that, after computing the nuclear matrix elements
${{\cal M}}_{p,n}$ the data provide constraints on the quantity $\cal Q$ of
Eq. (\ref{Rme.1}).
For the ground state to ground state transitions in spherically symmetric
nuclei the following integral representation is valid
\begin{equation}
{\cal M}_{p,n} = 4\pi \int j_0(p_e r) \Phi_\mu (r) \rho_{p,n} (r) r^2  dr
\label{V.1}
\end{equation}
where $j_0(x)$ the zero order spherical Bessel function and
$\rho_{p,n}$ the proton (p), neutron (n) nuclear density normalized to
the atomic number $Z$ and neutron number $N$ of the participating
nucleus, respectively. The space dependent part of the muon
wave function $\Phi_{\mu}$ is a spherically symmetric function which in our
calculations (see Sect. 4) was obtained by solving numerically the Shr\"ondinger
and Dirac equations with the Coulomb potential.

In defining $\cal Q$,  Eq. (\ref{Rme.1}), we introduced the ratio
\begin{eqnarray}\label{phi}
\phi = ({\cal M}_p - {\cal M}_n)/({\cal M}_p + {\cal M}_n)
\approx (A-2Z)/A,
\end{eqnarray}
where $A$ and $Z$ are the atomic weight and the total charge of the nucleus.
The quantity ${\cal Q}$ depends weakly on the nuclear parameters through the factor
$\phi$. In fact, the terms depending on $\phi$ are small since $\phi < 1$ (see
Table 1) and $G_S^{u}$,  $G_S^{d}$ as well as $G_V^{u}$, $G_V^{d}$ have the same
sign. In practice the nuclear dependence of ${\cal Q}$ can be neglected
and thus, ${\cal Q}$ coincides with the value of $2 \rho$ where $\rho$ is defined below.
It can be considered as a universal effective
\rp parameter measuring the \rp SUSY contribution to the $\mu-e$ conversion.
It also represents a suitable characteristic which allows
comparison of $\mu-e$ conversion experiments on different targets treating
the corresponding upper bounds on ${\cal Q}$ as their
sensitivities to the \rp SUSY signal.

For completeness, in Sect. 4 the limits for some non-SUSY
\cite{KosmasVergados96,R-cons} as well as $R_p$SUSY contributions to the 
$\mu^--e^-$ nuclear conversion (see in Fig. 2) are updated. 
The corresponding expression for $R_{\mu e^-}$ is written as \cite{KVF98}
        \begin{equation}
        R_{\mu e^-} = \rho\gamma,
        \label{III.2}
        \end{equation}
where $\rho$ is nearly independent of nuclear physics~\cite{mu-e-nucl} and 
contains the lepton flavor violating parameters corresponding to the 
contributions in Fig. 2. Thus, e.g. for photon-exchange mode $\rho$ is given by
\begin{equation}
\rho = (4\pi\alpha)^2\frac{\vert f_{M1}+f_{E0}\vert ^2+
\vert f_{E1}+f_{M0}\vert^2}{(G_Fm^2_{\mu})^2}
\label{rho}
\end{equation}
where the four electromagnetic form factors $f_{E0}$, $f_{E1}$, $f_{M0}$,
$f_{M1}$ are parametrized in a specific elementary model~\cite{mu-e-nucl}.

The factor $\gamma(A,Z)$ in Eq. (\ref{III.2}) accumulates about all the
nuclear structure dependence of the branching ratio $R_{\mu e^-}$.
Assuming that the total rate of the ordinary muon
capture is given by the Goulard-Primakoff function,
 $f_{GP}$, the nuclear structure factor $\gamma(A,Z)$ takes the form
        \begin{equation}
        \gamma(A,Z)\equiv\gamma = \frac{E_e p_e}{m_\mu^2}
        \frac{ M^2}{G^2 Z f_{GP}(A,Z)},
        \label{III.3}
        \end{equation}
where $G^2 \approx 6$. Thus, a non-trivial nuclear
structure dependence of the $\mu^-\to e^-$ conversion
branching ratio $R_{\mu e^-}$ is mainly concentrated
in the nuclear matrix elements $M^2$ \cite{KLV94}.
In the proton-neutron representation one can write down
        \begin{equation}
        M^2 = [ M_p+ Q \  M_n ]^2
        \label{ME2}
        \end{equation}
where $Q$ takes the values of Eq. (32) of Ref. [15a] and
$M_{p,n}$ are given by writing the matrix elements of Eq. (\ref{V.1}) in terms
of an effective muon wavefunction as
        \begin{equation}
        {\cal M}_{p,n} = \langle \Phi_{\mu} \rangle M_{p,n}
        \label{factoriz}
        \end{equation}
In our present approach the role of $\rho$ in Eq. (\ref{III.2}) is played 
by $\cal Q$, since ${\cal Q }= 2 \rho$, and the corresponding $\gamma$
function defined in Eq. (\ref{III.3}) for R-parity violating interactions,
$\gamma_{\rpm}$, is obtained from  
Eq. (\ref{III.3}) by putting $Q=1$ in Eq. (\ref{ME2}).
The separation of nuclear physics from the elementary particle parameters
is not complete but we have seen that $\phi$ is quite small. In any case
we present in Table 1 the values of $\phi$ for the various nuclear systems.

\section{Results and Discussion.}
The pure nuclear physics calculations needed for the $\mu-e$ conversion
studies refer to the integrals of Eq. (\ref{V.1}). The results of ${\cal M}_p$
and ${\cal M}_n$ for the currently
interesting nuclei $Al$, $Ti$ and $Pb$ are shown in Table 2. They have been
calculated using proton densities $\rho_{p}$ from the electron scattering
data \cite{Vries} and neutron densities $\rho_{n}$
from the analysis of pionic atom data \cite{Gibbs}.
We employed an analytic form for the muon wave function $\Phi_\mu(r)$ obtained by
solving the Schr\"odinger equation using the Coulomb potential produced
by the charge densities discussed above.
This way the finite size of a nucleus was taken into consideration.
Moreover, we included vacuum polarization corrections as in Ref. \cite{Chiang}.
In solving the Schr\"odinger equation we have used modern neural
networks techniques \cite{Lagaris} which give the wave function $\Phi_\mu (r)$
in the analytic form of a sum over sigmoid functions. Thus, in Eq. (\ref{V.1})
only a simple numerical integration is finally required.
To estimate the influence of the non-relativistic
approximation on the muon wave function $\Phi_\mu({\bf r})$, we have also
determined it by solving the Dirac equation. The results for
${\cal M}_{p,n}$
do not significantly differ from those of the Schr\"odinger picture.
In Table 2 we also show the muon binding energy $\epsilon_b$ (obtained
as output of the Dirac and Schr\"odinger solution) and the experimental values
for the total rate of the ordinary muon capture $\Gamma_{\mu c}$ taken from
Ref. \cite{Suzuki}.

Using the values of ${\cal M}_{p}$, ${\cal M}_{n}$ for a set of nuclei
throughout the periodic table we can estimate the nuclear
structure dependence of the $\mu-e$ conversion branching ratio, i.e. the
function $\gamma (A,Z)$ in Eq. (\ref{III.2}).
The results are quoted in Table 3. For comparison in this table we also
list the results of $\gamma_{\rpm}$ corresponding to the \rp SUSY mechanisms 
studied in the present paper and calculated as we stated in Sect. 3. 
These results can be exploited for setting constraints on
the quantities $\rho$ and $\cal Q$ corresponding to specific models
predicting the $\mu-e$ process.

In Table 4 we quote the upper bounds for the quantities $\rho$ and
$\cal Q$ corresponding to the mechanisms shown in Figs. 1,2.
These bounds were derived from the recent experimental upper bounds on
the branching ratio $R_{\mu e^-}$ for Ti and Pb targets given in
(\ref{Ti}) and (\ref{Pb}) and from the expected experimental sensitivity
(\ref{Al}) of the Brookhaven MECO experiment.
The limits on $\rho$ and ${\cal Q}$ quoted in Table 4 for $^{27}$Al
are improvements by about four orders of magnitude over the existing ones.

We should stress that limits on the quantities $\rho$ of Eq. (\ref{rho})
and ${\cal Q}$ of Eq. (\ref{Rme.1}), are the only constraints imposed by
the $\mu-e$ conversion on the underlying elementary particle physics.
One can extract upper limits on the individual lepton flavor violation
parameters (\rp couplings, effective scalar and vector couplings,
neutrino masses etc. \cite{Marci,KLV94,mu-e-nucl,Kos-Kov})
under certain assumptions like the commonly assumed
dominance of only one component of the $\mu-e$ conversion
Lagrangian which is equivalent to constrain one
parameter or product of the parameters at a time.
Using the upper limits for ${\cal Q}$ given in Table 3 we can derive
under the above assumptions the constraints on $\alpha^{(\tau)}_K$
of Eq. (\ref{Rme.1}) and the products of various \rp  parameters.
Thus, the bounds obtained for the scalar current couplings
$\alpha^{(0)}_{\pm S}$
in the $R$-parity violating Lagrangian for the $^{27}$Al target
\cite{SSK99} are $|\alpha^{(0)}_{\pm S}| < 7 \times 10^{-10}$.
The limit for $\alpha^{(0)}_{\pm S}$ obtained with the data for
the Ti target \cite{Schaaf} is $|\alpha^{(0)}_{\pm S}| < 1.1\times 10^{-7}$,
i.e. more than two orders of magnitude weaker than the limit of $^{27}$Al.

With these limits it is straightforward to derive constraints on
the parameters of the initial  Lagrangian (\ref{q-lev}). In Tables 5,6
we list the upper bounds on the products of the trilinear
\rp couplings
which we obtained from the experimental limit on $\mu-e$ conversion
in $^{48}$Ti and from the expected experimental sensitivity of MECO
detector using $^{27}$Al as a target material.
The corresponding constraints for $^{208}Pb$ are significantly weaker and
not presented here.
In Tables 5,6 the quantity $B$ denotes a scaling factor defined as
\begin{eqnarray}\label{scale}
B_{Ti} = (R^{exp}_{\mu e}/6.1 \times 10^{-13})^{1/2},\ \  \ \ \ 
B_{Al} = (R^{exp}_{\mu e}/2.0 \times 10^{-17})^{1/2},
\end{eqnarray}
which can be used for reconstructing the limits on the listed products
of the \rp parameters corresponding to the other
experimental upper limits on the branching ratio $R^{exp}_{\mu e}$.
In the 2nd column of Tables 5,6 we present previous limits existing
in the literature and taken from \cite{Huitu,reviews}.
As seen from Tables 5,6 the $\mu-e$-conversion limits on the products
$\lambda'\lambda'$ and $\lambda\lambda'$, except only
$\lambda'_{232}\lambda'_{132}$ and $\lambda'_{233}\lambda'_{133}$,
are significantly more stringent than those previously known in
the literature.
The two products $\lambda'\lambda'$ and $\lambda\lambda'$ are
less stringently constrained by the present experimental data on $^{48}$Ti
within the tree level non-photonic mechanism. Note that the corresponding
previous constraints on these products (2nd column of Table 5) were obtained
from the photonic 1-loop mechanism \cite{Huitu} of $\mu^--e^-$ conversion
which are better than existing in the literature non-$\mu^- \to e^-$
constraints on $\lambda'\lambda'$ and $\lambda\lambda'$.
The products in Table 6 are not constrained by this mechanism \cite{Huitu}.
However it constrains the products $\lambda\lambda$ not constrained by
the tree level \rp SUSY mechanism. At present the $\mu^-\to e^-$
constraints on  $\lambda\lambda$ within the 1-loop \rp SUSY mechanism
are weaker than those derived from the other processes.
As we have mentioned at the beginning, significant improvement on
these $\mu^- \to e^-$ limits is expected from the ongoing experiments
at PSI \cite{Schaaf} and even better from the MECO experiment at Brookhaven
\cite{Molzon}. This would make the  $\mu-e$ conversion constraints on
all the products of the \rp trilinear coupling attainable in the $\mu-e$
conversion at the tree and at the 1-loop levels better than those from
the other processes in all the cases.

Note that the last four limits for $\lambda'\lambda$ in
Table 6 originate from the contribution of the strange nucleon sea. These
limits are comparable to the other $\mu^- \to e^-$ constraints derived from
the valence quarks contributions.

Finally we put constraints on the products of the bilinear \rp  parameters
evaluated from the $\mu-e$ conversion in ${}^{48}$Ti as
\begin{eqnarray}
\label{bilin}
\lg\tilde\nu_1\rg\lg\tilde\nu_2\rg, \mu_1\mu_2,
\lg\tilde\nu_1\rg \mu_2, \lg\tilde\nu_2\rg \mu_1 \
\leq \left(80\mbox{MeV}\right)^2
\left(\frac{\tilde m}{100\mbox{GeV}}\right)^{2} B,
\end{eqnarray}
where $B$ is a scaling factor defined in Eq. (\ref{scale}).
These constraints are weaker than those derived in Ref. \cite{BFK}
from the Super-Kamiokande atmospheric neutrino data.

The limits in Tables 5,6 and in Eq. (\ref{bilin})
were extracted under the following assumptions. We
assumed that all the scalar masses in Eq. (\ref{coeff12})
are equal $\tilde m_{{u}L(n)}\approx \tilde m_{{d}L,R(n)}
\approx \tilde m_{\nu(n)}\approx \tilde m$ and also that
there is no significant compensation between the different
terms contributing to the ratio $R_{\mu e^-}$.
Note that the last assumption is, in practice,
equivalent to the other well known assumptions about the dominance of only
one parameter or product of the parameters at a time.
These assumptions are widely used in the literature for
derivation of constraints on the R-parity violating parameters.

Instead of considering only one specific combination of $\lambda$ and
$\lambda^\prime$ as dominant, one may attempt to consider all
combinations of the R-parity violating couplings. To this end one needs
constraints on the relative magnitudes of the R-parity violating
couplings. This can be accomplished in a fashion analogous to the theory
of textures for the Yukawa couplings. This theory, which expresses the
entries of the down fermion mass matrix in powers of a parameter
$\bar \epsilon = 0.23$, has been successful in describing the charged
fermion mass matrices (for review see \cite{lola:98} and references therein).
This can be extended to the R-parity violation
couplings themselves \cite{U1Rp,Leon-Riz}.
In fact using solution B of Ref. \cite{Leon-Riz}
we obtain
   \begin{equation}
\lambda^{\prime}_{112}=\lambda^{\prime}_{121}=\lambda^{\prime}_{212}=
\lambda^{\prime},
 \qquad \lambda_{1 2 1} = {\bar \epsilon}^2 \lambda^{\prime}
   \label{textures}
   \end{equation}
This way, $\mu - e$ conversion can be used to constrain the overall scale of
the R-parity violating interaction using phonomenologically acceptable textures 
\cite{Leon-Riz}.

\section{Summary and Conclusions.}

The transition matrix elements of the flavor violating $\mu^-- e^-$
conversion are of notable importance in computing accurately the corresponding
rates for each accessible channel of this exotic process.
Such calculations provide useful nuclear-physics inputs for the expected
new data from the PSI and MECO experiments to put more severe bounds on the
muon-number violating parameters
determining the effective currents
in various models that predict the exotic $\mu^-\to e^-$ process.

We developed a systematic approach which allows one to calculate
the $\mu^--e^-$ conversion rate in terms of the quark level Lagrangian
of any elementary model taking into account the effect of the nucleon
and nuclear structure.
Our conversion rate formula (\ref{Rme}) is valid for the interactions
with the (axial-)vector and (pseudo-)scalar quark and lepton currents,
as shown in Eq. (\ref{nucl1}). In the previous $\mu^- \to e^-$
calculations found
in the literature only the (axial-)vector currents were considered.

In the case of the R-parity violating interactions discussed here
we have investigated all the possible tree level contributions to the
$\mu-e$ conversion in nuclei.
We found new important contribution to $\mu^--e^-$ conversion
originating from the strange quark sea in the nucleon which
is comparable with the usual contribution of the valence $u,d$ quarks.

We introduced the quantities $\rho$ and $\cal Q$ defined
in Eqs. (\ref{rho}) and (\ref{Rme.1}) which can be associated with
theoretical sensitivities of a $\mu^--e^-$ conversion experiment
to the charged lepton flavor violating interactions discussed in the present
paper. These quantities are independent of a target material and, therefore,
might be helpful for comparison of searching potentials
of different $\mu-e$ conversion experiments and for planning future experiments.

>From the existing data on $R_{\mu e^-}$ in $^{48}$Ti and $^{208}$Pb
and the expected sensitivity of the designed MECO experiment \cite{Molzon}
we obtained stringent upper limits on the quantities $\rho$ and $\cal Q$.
Then we extracted the limits on the products of the trilinear \rp
parameters of the type $\lambda\lambda'$, $\lambda'\lambda'$
which are significantly more severe than those existing in the literature.

Let us conclude with the following important remark.
As was observed in Refs. \cite{Huitu,Kos-Kov},  if the ongoing
experiments at PSI \cite{Dohmen} and Brookhaven \cite{Molzon}
will have reached the quoted sensitivities in the branching ratio
$R_{\mu e^-}$ then the $\mu^- \to e^-$ constraints on all the products of
the \rp parameters $\lambda\lambda$, $\lambda'\lambda$,
$\lambda'\lambda'$ measurable in $\mu-e$ conversion
will become more stringent than those from any other processes.
This is especially important since no comparable improvements of
the other experiments probing these couplings is expected in the near future.

\vskip10mm
\noindent
{\bf Acknowledgments}\\

The research described in this publication was made possible
in part by the INTAS grant 93-1648-EXT and
Fondecyt (Chile) under grant 1000717. JDV would like to express
his appreciation to the Humboldt Foundation for their award and
thanks to the Institute of Theoretical Physics at University of
T\"ubingen for hospitality.
S.K. thanks V.A. Bednyakov and F. Simkovic for discussions.

\bigskip

\newpage
\begin{table}
\begin{center}
\begin{tabular}{rrrr}
\hline
\hline
 & & &  \\
 A & Z & $\phi$(A,Z) &  ${\tilde \phi}$(A,Z) \\
\hline
 & & &  \\
 12. &  6.  &  0.000  &  0.000  \\
 24. & 12.  &  0.014  &  0.000  \\
 27. & 13.  &  0.000  & -0.037  \\
 32. & 16.  &  0.023  &  0.000  \\
 40. & 20.  &  0.037  &  0.000  \\
 44. & 20.  & -0.063  & -0.091  \\
 48. & 22.  & -0.083  & -0.083  \\
 63. & 29.  & -0.056  & -0.079  \\
 90. & 40.  & -0.054  & -0.111  \\
112. & 48.  & -0.108  & -0.143  \\
208. & 82.  & -0.152  & -0.212  \\
238. & 92.  & -0.175  & -0.227  \\
\hline
\hline
\end{tabular}
\caption{ The variation of the quantity $\phi$ of Eq. (\ref{Rme.1})
through the periodic table. For comparison its approximate expression
$\tilde \phi \approx (A-2Z)/A$ is also shown.  }
\end{center}
\end{table}

\vspace{0.5 cm}

\begin{table}[ht]
\begin{center}
\begin{tabular}{rccccc}
\hline \hline
   &  &  &  &  &  \\
Nucleus & $|{\bf p}_e| \, (fm^{-1})$ & $\epsilon_b \, (MeV)$ &
$\Gamma_{\mu c} \, ( \times 10^{6} \, s^{-1})$ &
${\cal M}_p \, (fm^{-3/2})$ & ${\cal M}_n \, (fm^{-3/2})$  \\
\hline
   &  &  &  &  &  \\
$^{27}Al$  & 0.531 &  -0.470 &  0.71 & 0.047 & 0.045   \\
   &  &  &  &  &  \\
$^{48}Ti$  & 0.529 &  -1.264 &  2.60 & 0.104 & 0.127   \\
   &  &  &  &  &  \\
$^{208}Pb$ & 0.482 & -10.516 & 13.45 & 0.414 & 0.566   \\
\hline
\hline
\end{tabular}
\caption{Transition matrix elements (muon-nucleus overlap
integrals ${\cal M}_{p,n}$ of Eq. (\ref{V.1}))
evaluated by using the exact muon wave function obtained
via neural networks techniques. Other useful quantities
(see text) are also included.}
\end{center}
\end{table}

\vspace{0.5 cm}

\begin{table}
\begin{center}
\begin{tabular}{lccc}
\hline
\hline
 & & &  \\
Mechanism & \multicolumn{1}{c}{\large \bf $^{27}Al$}&
   \multicolumn{1}{c}{\large \bf $^{48}Ti$}&
   \multicolumn{1}{c}{\large \bf $^{208}Pb$} \\
\hline
 & & &  \\
 Photonic ($\gamma$)      &                   4.3
                     &              9.4
             &             17.3                   \\
W-boson exchange ($\gamma$)&                  34.2
                     &             25.3
             &             49.2                   \\
SUSY s-leptons ($\gamma$) &                  11.3
                     &             25.6
                     &             49.5                   \\
SUSY Z-exchange ($\gamma$) &                  27.6
                     &            110.6
                     &            236.2                   \\
\rp  SUSY ($\gamma_{\rpm}$)      &                  16.5
                     &             46.4
             &             96.9                   \\
\hline
\hline
\end{tabular}
\caption{
The nuclear physics part of the branching ratio $R_{\mu e}$,
i.e. values of the function $\gamma_{\rpm},\gamma$ of Eq. (\ref{III.3})
for the nuclear targets $^{27}$Al, $^{48}$Ti and $^{208}$Pb. }
\end{center}
\end{table}

\vspace{0.5 cm}

\begin{table}
\begin{center}
\begin{tabular}{lccc}
\hline
\hline
 & & &  \\
Mechanism & \multicolumn{1}{c}{\large \bf $^{27}Al$}&
   \multicolumn{1}{c}{\large \bf $^{48}Ti$}&
   \multicolumn{1}{c}{\large \bf $^{208}Pb$} \\
\hline
 & & &  \\
 Photonic       &      $\rho$ $\le$ 4.6 $\times 10^{-18}$
                     & $\rho$ $\le$ 8.2 $\times 10^{-14}$
             & $\rho$ $\le$ 3.2 $\times 10^{-12}$ \\
W-boson exchange&      $\rho$ $\le$ 5.8 $\times 10^{-19}$
                     & $\rho$ $\le$ 3.0 $\times 10^{-14}$
             & $\rho$ $\le$ 1.1 $\times 10^{-12}$ \\
SUSY sleptons  &      $\rho$ $\le$ 1.8 $\times 10^{-18}$
                     & $\rho$ $\le$ 3.0 $\times 10^{-14}$
                     & $\rho$ $\le$ 1.1 $\times 10^{-12}$ \\
SUSY Z-exchange &      $\rho$ $\le$ 7.3 $\times 10^{-19}$
                     & $\rho$ $\le$ 0.7 $\times 10^{-14}$
                     & $\rho$ $\le$ 0.2 $\times 10^{-12}$ \\
\rp  SUSY     &        ${\cal Q}$ $\leq$ 2.6 $\cdot 10^{-19}$
                     & ${\cal Q}$ $\leq$ 0.7 $\cdot 10^{-14}$
                     & ${\cal Q}$ $\leq$ 1.1 $\cdot 10^{-13}$ \\
\hline
\hline
\end{tabular}
\caption{Experimental limits on the elementary particle
sector of the exotic $\mu-e$ conversion. Quantities $\rho$
and ${\cal Q}$ are defined in of Eqs. (\ref{rho}) and Eq. (\ref{Rme.1}),
respectively.
The limits are extracted from the recent experimental data
for the nuclear targets $^{48}$Ti and $^{208}$Pb \protect\cite{Dohmen,Honec}
and from the expected sensitivity of the MECO experiment for the $^{27}$Al
target \protect\cite{Molzon}[see Eqs. (\ref{Ti})-(\ref{Al})]. }
\end{center}
\end{table}

\vspace{0.5 cm}

\begin{table}
\begin{center}
\begin{tabular}{|c|c|c|c|}
\hline \hline
Parameters & Previous limits & Present results& Expected results \\
& &$^{48}$Ti$(\mu-e)\cdot B_{Ti}$  & $^{27}$Al$(\mu-e)\cdot B_{Al}$ \\
\hline
&& &    \\
$|\lambda '_{211}\, \lambda'_{111}|$             &$4.4\cdot 10^{-6}$  &
$6.2\cdot 10^{-8}$& $4.0\cdot 10^{-10}$  \\%
$|\lambda '_{212}\, \lambda '_{112}|$            &$4.4\cdot 10^{-6}$  &
$1.7\cdot 10^{-8}$& $1.1\cdot 10^{-10}$  \\%
$|\lambda '_{213}\, \lambda '_{113}|$            &$4.4\cdot 10^{-6}$  &
$1.7\cdot 10^{-8}$& $1.1\cdot 10^{-10}$  \\%
$|\lambda '_{221}\, \lambda '_{111}|$            &$1.5\cdot 10^{-5}$  &

$7.6\cdot 10^{-8}$& $4.9\cdot 10^{-10}$   \\%
$|\lambda '_{222}\, \lambda '_{112}|$            &$1.5\cdot 10^{-5}$   &
$7.6\cdot 10^{-8}$& $4.9\cdot 10^{-10}$   \\%
$|\lambda '_{223}\, \lambda '_{113}|$            &$1.5\cdot 10^{-5}$   &
$7.6\cdot 10^{-8}$& $4.9\cdot 10^{-10}$  \\%
$|\lambda '_{231}\, \lambda '_{111}|$            &[$8.8\cdot 10^{-5}$]  &
$8.3\cdot 10^{-6}$& $5.3\cdot 10^{-8}$ \\%
$|\lambda '_{232}\, \lambda '_{112}|$            &$4.8\cdot 10^{-4}$  &
$8.3\cdot 10^{-6}$& $5.3\cdot 10^{-8}$ \\%
$|\lambda '_{233}\, \lambda '_{113}|$            &$4.8\cdot 10^{-4}$  &
$8.3\cdot 10^{-6}$& $5.3\cdot 10^{-8}$  \\%
$|\lambda '_{211}\, \lambda '_{121}|$            &$3.0\cdot 10^{-5}$   &
$7.6\cdot 10^{-8}$& $4.9\cdot 10^{-10}$   \\%
$|\lambda '_{212}\, \lambda '_{122}|$            &$3.0\cdot 10^{-5}$   &
$7.6\cdot 10^{-8}$& $4.9\cdot 10^{-10}$  \\%
$|\lambda '_{213}\, \lambda '_{123}|$            &$3.0\cdot 10^{-5}$   &
$7.6\cdot 10^{-8}$& $4.9\cdot 10^{-10}$  \\%
$|\lambda '_{221}\, \lambda '_{121}|$            &$8.0\cdot 10^{-6}$  &
$1.4\cdot 10^{-8}$& $9.0\cdot 10^{-11}$  \\%
$|\lambda '_{222}\, \lambda '_{122}|$            &$8.0\cdot 10^{-6}$  &
$3.3\cdot 10^{-7}$& $2.1\cdot 10^{-9}$  \\%
$|\lambda '_{223}\, \lambda '_{123}|$            &$8.0\cdot 10^{-6}$  &
$3.3\cdot 10^{-7}$& $2.1\cdot 10^{-9}$ \\%
$|\lambda '_{231}\, \lambda '_{121}|$            &$1.6\cdot 10^{-4}$  &
$3.7\cdot 10^{-5}$& $2.4\cdot 10^{-7}$   \\%
$|\lambda '_{232}\, \lambda '_{122}|$            &$1.6\cdot 10^{-4}$  &
$3.7\cdot 10^{-5}$& $2.4\cdot 10^{-7}$ \\%
$|\lambda '_{233}\, \lambda '_{123}|$            &$1.6\cdot 10^{-4}$  &
$3.7\cdot 10^{-5}$& $2.4\cdot 10^{-7}$ \\%
$|\lambda '_{211}\, \lambda '_{131}|$             &[$4.2\cdot 10^{-4}$]  &
$8.3\cdot 10^{-6}$& $5.3\cdot 10^{-8}$   \\%
$|\lambda '_{212}\, \lambda '_{132}|$            &$4.8\cdot 10^{-4}$  &
$8.3\cdot 10^{-6}$& $5.3\cdot 10^{-8}$   \\%
$|\lambda '_{213}\, \lambda '_{133}|$            &[$1.2\cdot 10^{-5}$] &
$8.3\cdot 10^{-6}$& $5.3\cdot 10^{-8}$  \\%
$|\lambda '_{221}\, \lambda '_{131}|$            &$1.6\cdot 10^{-4}$  &
$3.7\cdot 10^{-5}$& $2.4\cdot 10^{-7}$ \\%
$|\lambda '_{222}\, \lambda '_{132}|$            &$1.6\cdot 10^{-4}$  &
$3.7\cdot 10^{-5}$& $2.4\cdot 10^{-7}$ \\%
$|\lambda '_{223}\, \lambda '_{133}|$             &[$1.2\cdot 10^{-5}$]  &
$3.7\cdot 10^{-5}$& $2.4\cdot 10^{-7}$  \\%
$|\lambda '_{231}\, \lambda '_{131}|$            &$3.5\cdot 10^{-5}$  &
$1.3\cdot 10^{-8}$ & $8.3\cdot 10^{-11}$  \\%
$|\lambda '_{232}\, \lambda '_{132}|$            &$3.5\cdot 10^{-5}$  &
$4.0\cdot 10^{-3}$& $2.6\cdot 10^{-5}$   \\%
$|\lambda '_{233}\, \lambda '_{133}|$            &$3.5\cdot 10^{-5}$  &
$4.0\cdot 10^{-3}$& $2.6\cdot 10^{-5}$ \\%
\hline
\hline
\end{tabular}
\caption{Upper bounds on $\lambda'_{ijk}\lambda'_{lmn}$ derived
in the tree-level \rp SUSY mechanism from the SINDRUM II data on
the $\mu^--e^-$ conversion in ${}^{48}$Ti
[Eq. (\ref{Ti})] and from the expected
sensitivity [Eq. (\ref{Al})] of the MECO experiment at BNL with $^{27}$Al.
For comparison we present previous bounds derived in
the 1-loop \rp SUSY mechanism \protect\cite{Huitu}.
In brackets [ ] we show bounds from the other processes
(for references see Refs. \protect\cite{reviews,Huitu})
when they are more stringent then the 1-loop $\mu^--e^-$ conversion
bounds.
The limits are given for the scalar superpartner masses
$\tilde m = $ 100~GeV. The scaling factor $B$ is defined in Eq. (\ref{scale}).}
\end{center}
\end{table}

\newpage
\begin{table}
\begin{center}
\begin{tabular}{|c|c|c|c|}
\hline \hline
Parameters & Previous limits & Present results& Expected results \\
& &$^{48}$Ti$(\mu-e)\cdot B_{Ti}$  & $^{27}$Al$(\mu-e)\cdot B_{Al}$ \\
\hline
& & &   \\
$|\lambda '_{211}\, \lambda_{212}|$ &$4.5\cdot 10^{-3}$ &
$4.1\cdot 10^{-9}$& $2.6 \cdot 10^{-11}$\\
$|\lambda '_{311}\, \lambda_{312}|$ &$6.0\cdot 10^{-3}$ &
$4.1\cdot 10^{-9}$& $2.6 \cdot 10^{-11}$\\
$|\lambda '_{111}\, \lambda_{121}|$ &$1.5\cdot 10^{-5}$ &
$4.1\cdot 10^{-9}$& $2.6 \cdot 10^{-11}$\\
$|\lambda '_{311}\, \lambda_{321}|$ &$6.0\cdot 10^{-3}$ &
$4.1\cdot 10^{-9}$& $2.6 \cdot 10^{-11} $\\
$|\lambda '_{222}\, \lambda_{212}|$ &$9.0\cdot 10^{-3}$ &
$7.7\cdot 10^{-9}$& $5.0 \cdot 10^{-11} $\\
$|\lambda '_{322}\, \lambda_{312}|$ &$1.2\cdot 10^{-2}$ &
$7.7\cdot 10^{-9}$& $ 5.0 \cdot 10^{-11}$\\
$|\lambda '_{122}\, \lambda_{121}|$ &$1.0\cdot 10^{-3}$ &
$7.7\cdot 10^{-9}$& $ 5.0 \cdot 10^{-11}$\\
$|\lambda '_{322}\, \lambda_{321}|$ &$1.2\cdot 10^{-2}$ &
$7.7\cdot 10^{-9}$&  $5.0 \cdot 10^{-11}$\\
\hline
\hline
\end{tabular}
\caption{The same as in Table 5 but for the upper bounds  on
$\lambda'_{ijk}\lambda_{lmn}$. Previous limits are taken from
the other processes (for references see Refs. \protect\cite{reviews,Huitu}).
The 1-loop \rp SUSY mechanism \protect\cite{Huitu} does not constraint
these products of the \rp couplings. }
\end{center}
\end{table}

\end{document}